\documentclass{birkmult}
\usepackage{amssymb,amsmath,amsthm,mathrsfs,amsfonts}
\usepackage{latexsym}
\usepackage{physics}
\usepackage{calc}
\usepackage{leftidx}
\usepackage{multirow}
\usepackage{url}
\usepackage{graphics}
\usepackage{color}
\usepackage{graphicx}
\usepackage{float}
\usepackage{caption}
\usepackage{subcaption}
\usepackage{comment}
\usepackage{enumerate}
\usepackage{csquotes}
\renewcommand{\thesection}{\arabic{section}}
\renewcommand{\thesubsection}{\thesection.\arabic{subsection}}
\renewcommand{\thesubsubsection}{\thesubsection.\arabic{subsubsection}}
\numberwithin{equation}{section}

\usepackage[hidelinks]{hyperref}
\hypersetup{
    pdfnewwindow=true,
    plainpages=false,
}
\newcommand{\be}{\begin{equation}}
\newcommand{\ee}{\end{equation}}
\def\bal#1\eal{\begin{align}#1\end{align}}
\begin{document}
\title{Emergence of classicality from an inhomogeneous quantum universe}
\author{Adamantia Zampeli}
\address{Institute of Theoretical Physics, Faculty of Mathematics and Physics,
Charles University, \\
V Hole\v{s}ovi\v{c}k\'ach 2, 18000 Prague 8, Czech
Republic}
\email{azampeli@phys.uoa.gr}
\thanks{Prepared for the proceedings of the 2st Domoschool, Domodossola}
\date{\today}
\begin{abstract}
We give a short account of the quantisation of the Szekeres spacetime by considering the symmetries of a reduced action principle. This is an alternative approach than the one followed in the literature for the study of inhomogeneities, which is usually based on perturbations of an inhomogeneous field on the spacetime background. Here, we examine the emergence of classicality with an exact inhomogeneous solution. We check whether the two criteria for classicality are satisfied, that is the correlations on the phase space and decoherence. We verify that these two properties indeed hold, thus the classical behaviour emerges from our considerations. We comment on the connection between the emergence of an inhomogeneous spacetime and the current cosmological observations of a highly homogeneous universe at large scales. 
\end{abstract}
\maketitle
\section{Introduction}
More than a century now, we know that the world is fundamentally quantum mechanical; yet our everyday experience fools us with a classical world. How do these two pictures reconcile? Indeed, if one accepts the thesis that the world is fundamentally quantum mechanical, the natural question to ask is how classical world we experience in our everyday life emerges from the quantum structures. There must be a mechanism for this to happen; but more importantly, what are the requirements for a system to be considered classical? Here, we will adopt the position that this transition from quantum to classical happens through the mechanism of decoherence, which destroys the interference terms between different quantum states. Usually, this happens through the interaction of the system with an environment. In quantum cosmology, on which we focus our considerations here, the universe is a closed system and the role of the environment is played by inhomogeneous degrees of freedom acting as perturbations in an overall homogeneous background (e.g. \cite{Halliwell:1989vw,Barvinsky:1998cq}). We are interested to explore a different path, by starting with a genuinely inhomogeneous spacetime and investigating the emergence of classicality due to the presence of symmetries.

Our starting point is the Szekeres spacetime, which is a dust type D exact solution of general relativity and has classically attracted the interest as alternative to Friedman-Lemaitre-Robertson-Walker (FLRW) description of the universe, but also of the Bianchi I (Kasner). The first currently serves as the ``standard" cosmological model, while the latter as a model for the dynamics close to the singularity \cite{Belinsky:1970ew}. It is therefore clear that the class of models described by the Szekeres spacetimes can provide useful information for many properties of physically interesting models and possible new effects which might appear due to quantum gravity. 

There are two criteria we consider to examine the emergence of classicality: i) Hartle's criterion which states that predictions in quantum cosmology are possible when there is a peak on the configuration space, which accordingly indicates a correlation between conjugate momenta on the phase space \cite{Hartle:1986gn,10.2307/2214880} and ii) decoherence, which is quantified in the condition that the sum of the non-diagonal terms of the reduced density matrix are much smaller than the sum of its diagonal terms \cite{Halliwell:1989vw,Dowker:1992es}. To this end, we define the reduced density matrix by tracing out the constant of motion related to the classical symmetry in question. 

In the following sections, we first summarise previous results regarding the classical symmetries of a reduced Lagrangian for the Szekeres spacetime and the solution of the quantum equations. Then, we proceed to check whether the first, but mainly the second criterion holds for a reduced density matrix defined as described above. In the last section, we discuss the results and the connection with the current cosmological observations about the homogeneity of the universe.
\section{Preliminary results}
The general spacetime element for the Szekeres solution is \cite{Szekeres:1974ct}
\be
ds^2 = -dt^2 + e^{2A (t,x,y,z)} dx^2 + e^{2B(t,z,y,z)} (dy^2 +dz^2)
\ee
where the functions $A(t,x,y,z), \ B(t,x,y,z)$ can be specified by the solution of the Einstein equations with energy-momentum tensor of the dust,
\be
G_{\mu\nu} = T_{\mu\nu}^{(D)}
\ee
Instead of the metric variables, we choose the physical variables, since we can take advantage of the fact that in these solutions the two components of the electric part of the Weyl tensor and the two components of the shear for the observer denoted by a time-like $4-$vector $u^{\mu}$ are equal respectively. In these variables, the evolution equations take the form \cite{Paliathanasis:2017wli}
\begin{subequations}
\label{szeksys}
\begin{align}
&\dot{\rho}+\theta\rho =0,~  \label{ss.01} \\
&\dot{\theta}+\frac{\theta^{2}}{3}+6\sigma^{2}+\frac{1}{2}\rho =0,
\label{ss.02} \\
&\dot{\sigma}-\sigma^{2}+\frac{2}{3}\theta\sigma+E =0,  \label{ss.03} \\
&\dot{E}+3E\sigma+\theta E+\frac{1}{2}\rho\sigma =0,  \label{ss.04}
\end{align}
\end{subequations}
where $\dot{}=u^{\mu}\nabla_{\mu}$ and the energy density is defined as $\rho=T^{\mu\nu }u_{\mu}u_{\nu}$. The constraint equation is
\begin{equation}
\frac{\theta^{2}}{3}-3\sigma^{2}+\frac{^{\left(  3\right)  }R}{2}=\rho,
\end{equation}

We can find a second-order differential system by solving equation \eqref{ss.01} with respect to $\theta$ and equation \eqref{ss.04} with respect to $\sigma$ and replacing the results to the other two. The new system, which we omit to write can be further simplified through the transformation
\begin{equation}\label{transf}
\rho=\frac{6}{u^3(1-\frac{v}{u})}, \ E=\frac{v}{u^{4}\left( \frac{v}{u}-1\right) }
\end{equation}
thus taking the simplified form
\bal\label{szeksys}
&\ddot{v} - \frac{2v}{u^3}=0, \\
&\ddot{u} + \frac{1}{u^2} = 0 \label{e-l2}
\eal
These correspondingly can be seen as the Euler-Lagrange equations of the following Lagrangian function
\be
L= \dot{u} \dot{v} -\frac{v}{u^2}
\ee
It is interesting to note that the initial spacetime, despite the degeneracy between the different components of the Weyl tensor and the shear has no Killing vector field and it can only be said that these solutions are locally axisymmetric \cite{Bruni:1994nf}. However, this reduced Lagrangian possess generalised symmetries which facilitates the quantisation of this system \cite{Paliathanasis:2018ixu}. The symmetries of this Lagrangian are
\begin{subequations}\label{symm_sys}
\bal
&h= \dot{u} \dot{v} -\frac{v}{u^2}, \\
&I_0 = \dot{u}^2 -\frac{2}{u}
\eal
\end{subequations}
and in the appendix \ref{killingtensors} it is shown that their presence is due to the existence of two trivial Killing tensor fields on the confugration space of the $(u,v)$ variables. It is clear that the first equation is the Hamiltonian function, thus it plays the role of ``energy'' of the reduced system. The stability analysis of this system \cite{Paliathanasis:2017wli} showed that there are two exact solutions when $h=0$ and $I_0=0$ of the form $u (t) =u_0 z^{-1}, v(t) = v_0 t^{2/3}$ and $u(t) =u_0 t^{2/3}, v(t)= v_0 t^{2/3}$. The first solution corresponds to an unstable critical point for the dynamical system \eqref{szeksys} while the latter to a stable one \cite{Paliathanasis:2017wli}. 
\section{Quantum dynamics and classical emergence}
We now turn to the quantum dynamics which arises by turning to quantum operators the fundamental variables on the phase space and the classical observables to self-adjoint operators. Then, we find the following eigen-equations \cite{Paliathanasis:2018ixu}
\bal
&\left( -\partial _{uv}+\frac{v}{u^{2}}\right) \Psi =h\Psi ,  \label{sss.31} \\
&\left( \partial _{vv}+\frac{2}{u}\right) \Psi =-I_{0}\Psi ,  \label{sss.32}
\eal
which are the quantum analogues of \eqref{symm_sys}. We note in passing that, contrary to what happens for gravitational systems, which are constrained due to the presence of the arbitrary functions (the lapse function and the shift vector), here the dynamics of the reduced system is not constrained. We now limit ourselves to the case $h=0$, in which the wave function takes the form
\begin{equation}
\Psi \left( I_{0},u,v\right) =\frac{\sqrt{u}}{\sqrt{2+I_{0}u}}\left(
\Psi _{1}\cos \left( \sqrt{\frac{2+I_{0}u}{u}}v\right) +\Psi _{2}\sin \left(
\sqrt{\frac{2+I_{0}u}{u}}v\right) \right) .
\end{equation} 
If we select the constants such that $\Psi_2 = i \Psi_1 = C$, the wave function is written in polar form as
\be\label{polar_form}
\Psi (I_0, u,v) = \frac{C \sqrt{u}}{\sqrt{2+I_{0}u}} \exp \left( i \sqrt{\frac{2+I_{0}u}{u}} v \right)
\ee
where $C$ is a constant. In \cite{Paliathanasis:2018ixu}, we performed the Bohmian analysis for this wave function as well as for general values of the constant $h$. It was shown that the quantum potential, which is given by 
\be
\mathcal{Q} (q^i) = -\frac{\square \Omega (q^i)}{2 \Omega (q^i)}. 
\ee
where $\Omega (q^i)$ is the amplitude of the wave function \eqref{polar_form}, $q^i$ are the variables of the configuration space, in this case $(u,v)$ and $\square$ the Laplacian for this space, becomes zero. Since the quantum potential appears in the Hamilton-Jacobi equation as an additional term arising due to quantum effects, its value becoming zero means that the classical dynamics emerges from these quantum solutions. This can also be attributed to the fact that the variables $q_i = u,v$ and their conjugate momenta which are defined as $p_i = \nabla_i S$ are highly correlated. Indeed, following the analysis of \cite{Halliwell:1987aa}, where it was shown that for the case of WKB-type wave functions strong correlations between the variables and their conjugate momenta on the phase space lead to strong peaks of the wave function and to classicality. We can conclude that the first criterion for the emergence of classicality as introduced by Hartle is satisfied \cite{Hartle:1986gn,10.2307/2214880}.

We are now interested to check whether the second criterion holds, which is decoherence; this is the destruction of the interference terms between different systems due to correlations \cite{Zurek_2006} and happens due to the interaction between subsystems. One plays the role of environment, which has infinite degrees of freedom and is of no interest in the analysis. Therefore it is traced out, keeping only the degrees of freedom of the system under physical interest. In cosmology, the environment is usually inhomogeneous degrees of freedom of some scalar field. In our case, though, we are interested to examine decoherence in relation to the existence of a symmetry. The induction of decoherence due to symmetries has been discussed before elsewhere, e.g. \cite{Giulini:1994zh,Giulini:1997nz}. In order to quantify this effect, we will define the reduced density matrix as
\be
\rho^{red} (u_i,v_j,u_k,v_l) = \ket{\Psi (u_i,v_j)} \bra{\Psi (u_k,v_l)} =\int_0^\alpha DI_0 \Psi^* (u_i,v_j,I_0) \Psi (u_k,v_l,I_0)
\ee
i.e. by tracing out the symmetry constant $I_0$ and $i,j=1,2$. If we insert the polar form of the wave function it becomes
\bal\label{nondiag_red}
\rho^{red}_{ijkl} = \int_0^\alpha dI_0 \Omega^*(u_i, I_0) \Omega (u_k, I_0) e^{-i (S (u_i,v_j,I_0)- S (u_k,v_l,I_0))}
\eal
The condition for decoherence is that the sum of the real part of the non-diagonal elements of this matrix should be much smaller than the sum of the diagonal elements \cite{Dowker:1992es}
\be\label{decoherence_cond}
\abs{\sum_{i\neq j} \Re\rho^{red} (u_i,u_j)} < \epsilon \sum_{i=j} \rho^{red} (u_i,u_j).
\ee
The diagonal elements are the ones with 
\be\label{diagonal_elem}
\rho^{red}_{ijij} = \int_0^\alpha d I_0 \Psi^* (u_i,v_j) \Psi (u_i,v_j) = \int_0^\alpha dI_0 \abs{\Omega (u_i, I_0)}^2 
\ee
which, after substituting the explicit form of the solution become
\be
\rho^{red}_{diag}=\int_0^\alpha d I_0  \frac{\abs{C}^2 u}{2 + I_0 u} = \abs{C}^2 \ln (1+\frac{\alpha u}{2})
\ee
and depend only on $u$. For the non-diagonal elements, we are interested in the behaviour of their real part. These are given by $i \neq k$ and/or $j \neq l$. 
Their real part is given by
\be
\Re\rho^{red} (u_i,v_j,u_k,v_l,I_0) = \int_0^\alpha d I_0 \Omega^*(u_i,I_0) \Omega  (u_k,I_0)\left( \cos (S_{ij} - S_{kl}) \right)
\ee
This expression is always bounded, i.e always satisfies the relationship
\be
\abs{\Re\rho^{red} (u_i,v_j,u_k,v_l,I_0)} \leq \int_0^\alpha d I_0 \Omega^*(u_i,I_0) \Omega  (u_k,I_0)
\ee
with the equality holding for the diagonal elements when $S_{ij} = S_{kl}$. The right-hand side can be calculated and it is equal to
\be
\abs{C}^2 \ln \left(\frac{u_i \left(\alpha u_j+1\right)+\sqrt{u_i u_j \left(\alpha u_i+2\right) \left(\alpha u_j+2\right)}+u_j}{2 \sqrt{u_i u_j}+u_i+u_j}\right)
\ee
from which we recover the relation \eqref{diagonal_elem} for $i=j$. The relation we wish to show that holds for all the range of values of $u$ is the sum of the corresponding term i.e.
\be
\abs{2 \Re \rho^{red} (u_1,u_2)} < \epsilon ( \rho^{red} (u_1, u_1) + \rho^{red} (u_2,u_2) )
\ee
since $\Re \rho^{red} (u_1,u_2) = Re \rho^{red} (u_2,u_1/)$ which is written in our case as 
{\small
\bal
2\ln\left( \frac{\sqrt{u_i u_j \left(\alpha u_i+2\right) \left(k u_j+2\right)}+u_i \left(\alpha u_j+1\right)+u_j}{2 \sqrt{u_i u_j}+u_i+u_j} \right) <
\epsilon \left(\ln \left(\frac{\alpha u_i}{2}+1\right)+\ln \left(\frac{\alpha u_j}{2}+1\right)\right)
\eal}
This relation is true for every $\alpha>0$ and positive values of the configuration variables, which is of our interest, therefore we do have decoherence for this case induced by the presence of symmetry.
\section{Discussion}
We studied the quantum solution of an inhomogeneous gravitational model which is an exact classical solution. We showed that the presence of symmetry can satisfy two criteria for the emergence of classicality for the particular case of $h=0$. Instead of separating our system to environment and subsystem, we traced out over the classical constant of motion $I_0$. We examined the possibility that the interference terms are destroyed due to the existence of symmetries and we found that this can indeed be the case. It is a known fact that symmetries can lead to decoherence and this can also be manifest formally through the existence of superselection rules . These are rules prohibiting the existence of pure states which are superpositions of states belong to different coherent subspaces of the Hilbert space \cite{Giulini:1997nz}. 

These considerations strengthen the results in \cite{Paliathanasis:2018ixu} and give further motivation to study possible quantum effects at the low-energy limit coming from inhomogeneous spacetimes.

\appendix
\section{The Killing tensors of the Lagrangian}\label{killingtensors}

The metric on the configuration space is
\be
G_{\mu\nu}=
\begin{pmatrix}
0 & 1 \\
1 & 0%
\end{pmatrix}
\ee
The Killing fields are
\bal
&\xi_1 = \partial_u, \ \xi_2 = \partial_v, \ \xi_3 = u \partial_u - v \partial_v
\eal
The (trivial) Killing tensors constructed by these Killing vector fields are found by the relation
\be
K = \frac{1}{2} \left( \xi_i \otimes \xi_j + \xi_j \otimes \xi_i \right)
\ee
and have the form
\be
K_1 =
\begin{pmatrix}
1 & 0 \\
0 & 0%
\end{pmatrix}
\ee
and
\be
K_2 =
\begin{pmatrix}
0 & 1 \\
1 & 0%
\end{pmatrix}
\ee
The conserved quantities are given by $K_i = K_i^{\mu\nu} p_\mu p_\nu$ and one can see that $K_1$ corresponds to the equation \eqref{e-l2}, while $K_2$ to the energy, thus can be associated with the constants of motion considered in the main text as $K_1 \rightarrow I_0$ and $K_2 \rightarrow h$.
\subsection*{Acknowledgments}
I would like to thank the organisers of the 2nd Domoschool for the high level of lectures and their kind hospitality. During the school, I was benefited from discussions with Profs. Sergio Cacciatori, Vittorio Gorini and Alexander Kamenshchik. I also thank Drs. Andronikos Paliathanasis, Georgios Pavlou and Otakar Svitek for suggestions and corrections of the manuscript. Finally, I acknowledge financial support from the Albert Einstein Center.

\bibliographystyle{jhep}
\bibliography{szekeres}

\providecommand{\href}[2]{#2}\begingroup\raggedright\begin{thebibliography}{10}

\bibitem{Halliwell:1989vw}
J.~J. Halliwell, \emph{{Decoherence in Quantum Cosmology}},
  \href{http://dx.doi.org/10.1103/PhysRevD.39.2912}{\emph{Phys. Rev.} {\bf D39}
  (1989) 2912}.

\bibitem{Barvinsky:1998cq}
A.~O. Barvinsky, A.~{\relax Yu}. Kamenshchik, C.~Kiefer and I.~V. Mishakov,
  \emph{{Decoherence in quantum cosmology at the onset of inflation}},
  \href{http://dx.doi.org/10.1016/S0550-3213(99)00208-4}{\emph{Nucl. Phys.}
  {\bf B551} (1999) 374--396}, [\href{http://arxiv.org/abs/gr-qc/9812043}{{\tt
  gr-qc/9812043}}].

\bibitem{Belinsky:1970ew}
V.~A. Belinsky, I.~M. Khalatnikov and E.~M. Lifshitz, \emph{{Oscillatory
  approach to a singular point in the relativistic cosmology}},
  \href{http://dx.doi.org/10.1080/00018737000101171}{\emph{Adv. Phys.} {\bf 19}
  (1970) 525--573}.

\bibitem{Hartle:1986gn}
J.~B. Hartle, \emph{{Prediction in Quantum Cosmology}},
  \href{http://dx.doi.org/10.1007/978-1-4613-1897-2_12}{\emph{NATO Sci. Ser. B}
  {\bf 156} (1987) 329--360}.

\bibitem{10.2307/2214880}
R.~Geroch, \emph{The everett interpretation}, {\emph{No{\^u}s} {\bf 18} (1984)
  617--633}.

\bibitem{Dowker:1992es}
H.~F. Dowker and J.~J. Halliwell, \emph{{The Quantum mechanics of history: The
  Decoherence functional in quantum mechanics}},
  \href{http://dx.doi.org/10.1103/PhysRevD.46.1580}{\emph{Phys. Rev.} {\bf D46}
  (1992) 1580--1609}.

\bibitem{Szekeres:1974ct}
P.~Szekeres, \emph{{A Class of Inhomogeneous Cosmological Models}},
  \href{http://dx.doi.org/10.1007/BF01608547}{\emph{Commun. Math. Phys.} {\bf
  41} (1975) 55}.

\bibitem{Paliathanasis:2017wli}
A.~Paliathanasis and P.~G.~L. Leach, \emph{{Symmetries and Singularities of the
  Szekeres System}},
  \href{http://dx.doi.org/10.1016/j.physleta.2017.02.009}{\emph{Phys. Lett.}
  {\bf A381} (2017) 1277--1280}, [\href{http://arxiv.org/abs/1702.01593}{{\tt
  1702.01593}}].

\bibitem{Bruni:1994nf}
M.~Bruni, S.~Matarrese and O.~Pantano, \emph{{Dynamics of silent universes}},
  \href{http://dx.doi.org/10.1086/175755}{\emph{Astrophys. J.} {\bf 445} (1995)
  958--977}, [\href{http://arxiv.org/abs/astro-ph/9406068}{{\tt
  astro-ph/9406068}}].

\bibitem{Paliathanasis:2018ixu}
A.~Paliathanasis, A.~Zampeli, T.~Christodoulakis and M.~T. Mustafa,
  \emph{{Quantization of the Szekeres System}},
  \href{http://dx.doi.org/10.1088/1361-6382/aac227}{\emph{Class. Quant. Grav.}
  {\bf 35} (2018) 125005}, [\href{http://arxiv.org/abs/1801.01276}{{\tt
  1801.01276}}].

\bibitem{Halliwell:1987aa}
J.~J. Halliwell, \emph{Correlations in the wave function of the universe},
  \href{http://dx.doi.org/10.1103/PhysRevD.36.3626}{\emph{Phys. Rev.} {\bf D36}
  (1987) 3626--3640}.

\bibitem{Zurek_2006}
W.~H. Zurek, \emph{Decoherence and the transition from quantum to classical ---
  revisited},
  \href{http://dx.doi.org/10.1007/978-3-7643-7808-0_1}{\emph{Quantum
  Decoherence} (2006) 1--31}.

\bibitem{Giulini:1994zh}
D.~Giulini, C.~Kiefer and H.~D. Zeh, \emph{{Symmetries, superselection rules,
  and decoherence}},
  \href{http://dx.doi.org/10.1016/0375-9601(95)00128-P}{\emph{Phys. Lett.} {\bf
  A199} (1995) 291--298}, [\href{http://arxiv.org/abs/gr-qc/9410029}{{\tt
  gr-qc/9410029}}].

\bibitem{Giulini:1997nz}
D.~Giulini, \emph{Superselection rules and symmetries},
  \href{http://dx.doi.org/10.1007/978-3-662-03263-3_6}{\emph{Decoherence and
  the Appearance of a Classical World in Quantum Theory} (1996) 187--222}.

\end{thebibliography}\endgroup
\end{document}